\documentclass[12pt]{article}

\usepackage{a4wide}
\usepackage{epsfig}

\begin{document}

\title{Rheology of a confined granular material}

\author{Guillaume Ovarlez, Evelyne Kolb and Eric Cl\'{e}ment \\
Laboratoire des Milieux D\'{e}sordonn\'{e}s et H\'{e}t\'{e}rog\`{e}nes \\
UMR7603 - Universit\'{e} Pierre et Marie Curie - Bo\^{\i }te 86 \\
4, Place Jussieu, 75005\ Paris, France
} 

\date{\today} 

\maketitle

\begin{abstract} 
We study the rheology of a granular material 
slowly driven in a confined geometry. The motion is characterized by a steady
sliding with a resistance force increasing with the driving velocity and the
surrounding relative humidity. For lower driving velocities a transition to
stick-slip motion occurs, exhibiting a blocking enhancement whith decreasing
velocity. We propose a model to explain this behavior pointing out the leading
role of friction properties between the grains and the container's boundary. 
\end{abstract}

\vspace{1cm}

Dynamics of granular materials lacks of an established unified
picture. A great diversity of mechanical and rheological behaviors were
reported depending on whether they are vibrated, slowly sheared, avalanching
on a surface, flowing in a hopper or falling in a chute\cite{PDM}. For dense
granular assemblies, experiments reveal strain localization \cite{Howell99}, 
\cite{Mueth2000}, non local rheological properties\cite{Pouliquen99} and
aging phenomena\cite{Gollub97},\cite{Boquet98}. This complex phenomenology
could possibly be due to the presence, at a mesoscopic level, of a
disordered contact force network with unraveled mechanical properties \cite
{Howell99},\cite{Radjai98}. Moreover, dramatic effects on the mechanical
strength induced by slight changes in compaction were reported\cite{Howell99}%
, \cite{Horwarth96},\cite{Vanel99}. Another source of difficulty is the
dynamical contribution of contact \ forces and there is, so far, only few
studies on the macroscopic emergence of these aging properties due to the
slow plastic deformation of contacts and the influence of surrounding
humidity \cite{Boquet98},\cite{Crassous99},\cite{Dieterich84},\cite{Halsey98}%
.

Dynamical behavior of slowly driven granular materials was investigated
recently by different groups both in compression and/or in shear experiments 
\cite{Mueth2000},\cite{Gollub97},\cite{Horwarth96},\cite{Albert99},\cite
{LaKolb99}. Here, we investigate the rheology of a granular assembly
confined in a cylindrical column and pushed vertically from the bottom. The
resistance to vertical motion as well as the blocking/unblocking
transitions, reveal a phenomenology possibly shared by many confined
granular assemblies, as for example, gouge sheared between two faults\cite
{Marone98}, pipe flows, compaction under stress or dense granular paste
extrusion. A previous investigation of the same display but in 2D\cite
{LaKolb99} has already shown a rich phenomenology. For 3D granular
assemblies, we get similar behavior but in this letter, we choose to report
only on the most simple dynamical situation involving low friction grains
confined in a column with rather frictional walls. Here, we have a weak
coupling with dilatancy effects due to shearing at the walls and therefore,
relations between the granular nature of the bulk (i.e. the stress
redirection) and the solid friction properties at the walls are the most
clearly revealed.

The grains are dry, non cohesive and monodisperse steel beads of diameter $%
d=1.58$ $mm$ piled into a vertical cylinder in duralumin of diameter $D=36$ $%
mm$. The column is closed at the bottom by a movable piston avoiding contact
with the column (diameter mismatch is $0.5$ $mm$). A force probe of
stiffness $k=40000$ $N.m^{-1}$ is located under the piston and is pushed at
a constant driving velocity $V$ (between $5$ $nm.s^{-1}$ and $100$ $\mu
m{}.s^{-1}$) via a stepping motor (see inset of fig.1a). The resistance
force $F$ encountered by the piston is measured as a function of time. We
monitor also the relative humidity ($RH$) and the surrounding temperature.
The central question here is an attempt to estimate the relative influence
of the bulk mechanical properties with respect to frictional properties of
the walls. To address directly this issue we built a special device (the
slider) designed to apply a constant normal load ($F_{N}=2$ $N$) on three
steel beads sliding vertically on the cylinder's wall (see inset of fig.3a).
Then, the dynamical evolution of the resistance force encountered by the
piston pushing the grains is compared with the slider's friction resistance
driven in the same conditions. We observe two distinct regimes(fig.1a): for
high driving velocities, the motion is characterized by a steady-sliding and
a constant pushing force; for low velocities, the system undergoes a dynamic
instability and then a stick-slip motion occurs. The transition between
these behaviors is similar to the inertial regime of Heslot et al. \cite
{Heslot94} and details will be reported elsewhere.

For a vertically pushed granular assembly, the driving force exerted by the
piston is screened. To evaluate this effect, the mean resistance force is
measured as a function of the packing height (see fig.1b). For this dataset
the driving velocity $V$\ corresponds to a steady and continuous sliding of
the grains. The resistance force $F$ increases very rapidly with the
column's height $H$. This strong resistance to motion is due to the
horizontal redirection of stresses in association with solid friction at the
side walls. Following the standard Janssen's screening picture\cite{Vanel99}%
, \cite{Nedderman92}, the force $F$ exerted by the grains on the piston can
be modelled by the relation:

\begin{equation}
F_{\epsilon }=\varrho g\lambda \pi R^{2}\times \epsilon (\exp (\epsilon 
\frac{H}{\lambda })-1)  \label{1}
\end{equation}

where $\varrho $ is the mass density of the granular material, $R$ is the
cylinder radius and $g$ the acceleration of gravity. The length $\lambda
=R/2K\mu $ is the so-called effective screening length, where $K$ is the
Janssen's parameter rendering the average horizontal redirection of vertical
stresses and $\mu $ can either be the dynamic or the static coefficient of
friction of beads at the cylinder's wall. When $\epsilon =+1,$ friction is
fully mobilized downwards (our pushing experiment) and when $\epsilon =-1,$
friction is fully mobilized upwards. It is easily seen from (\ref{1}) that
when $\epsilon =+1$, any slight change in $\mu $ or $K$ is exponentially
amplified with a drastic influence on the pushing force $F$. In the case of
steel beads, we found that, starting from a dense or a loose packing, the
final average steady state compacity $\stackrel{\_}{\nu }$ does not change ;
we have $\stackrel{\_}{\nu }\approx 62.5\%$ for all velocities and relative
humidities $RH$ tested. In the steady state regime, the experimental data
obtained for a given pushing velocity $V$ can be fitted with relation (\ref
{1}) by adjusting only one parameter i.e. $p_{+1}=K\times \mu _{d}$ where $%
\mu _{d}$ is the dynamic coefficient of friction at velocity $V$. For the
relative humidity $RH=42\%$, we obtain $p_{+1}=0.140\pm 0.001$ at $16$ $\mu
m.s^{-1}$ and $p_{+1}=0.146\pm 0.001$ at $V_{up}=100$ $\mu m.s^{-1}$. As a
check of consistency, we performed the following dynamical experiment.
First, the granular column is pushed upwards in order to mobilize the
friction forces downwards and to reach the steady state compacity. Starting
from this situation, the friction forces are reversed at the walls by moving
the piston downwards at a constant velocity $V_{down}=16\mu m.s^{-1}$.
Following relation (\ref{1}), this procedure should imply a change of $%
\epsilon \ $from $1$ to $-1$, and consequently, the dynamical force on the
piston should decrease from $F_{+1}$ to $F_{-1}$. In fig.1b the pushing
force $F_{-1}$ is measured for different packing heights $H$. Injecting the
preceding value of $p_{+1}(16$ $\mu m.s^{-1})$ into (\ref{1}) with $\epsilon
=-1,$\ we check on fig1b, that the theoretical expectation agrees quite well
with the experimental data of $F_{-1}$ versus $H$. Note that in a previous
study it was found that the Janssen's picture has a general tendency to
underestimate the stress below a granular column \cite{Vanel99}. But in the
present situation, with low friction steel beads, this model though
elementary, seems a good base for analysis. Nevertheless, a question is
still that the fitting parameters $p=K\times \mu $ extracted from the model
are unable to sort between what comes out from wall-bead interactions ($\mu $%
) and what comes out from bulk properties ($K$). Actually, from series of
static Janssen experiments we extracted $K\mu _{s}$ . Independently, the
static coefficient of friction $\mu _{s}$ of our steel beads on duralumin
was measured (in the short time limit) using the sliding angle of a three
beads tripod. We get $\mu _{s}=0.170\pm 0.005$ and $K=1.08\pm 0.05$ is
extracted from this procedure. This $K$ value is consistent with previous
measurements done on a granular column at this compacity\cite{Vanel99}. Note
that, if we tentatively assume a constant value for $K$ in static and
dynamic experiments, the dynamical coefficient $\mu _{d}$\ can be extracted
from the measured values of $p_{+1}$. For instance, for $RH=42\%$ we extract
the values, $\mu _{d}(16\mu m.s^{-1})=0.130\pm 0.005$ and $\mu _{d}(100\mu
m.s^{-1})=0.135\pm 0.005$. This would imply a slight increase of the
bead/wall friction with the driving velocity. This result is going to be
directly tested in the following, using the three beads slider device.

For a given height $H$ of beads ($M=380g$ i.e. $H=4.3R$), we study
extensively how the pushing force depends on the driving velocity $V$ and on
the surrounding relative humidity ($RH$). We worked in the range $%
35\%<RH<75\%$, and also in dry air ($RH<3\%$). Note that except for the dry
situation, we did not have a mean of regulation of this last parameter ($RH$%
) but we record its values around the experimental set-up. All data are
shown on a same series for similar humidity values (within $3\%$). As it was
already mentioned, the motion is characterized by a steady sliding above a
critical velocity (fig.1a). The mean force level in this regime increases
slowly with velocity but surprisingly strongly with $RH$ (fig.2a). For
example at $V=100$ $\mu m.s^{-1}$, the resistance force is raised by $35\%$
for a change of $RH$ from $53\%$ to $72\%$. Now we perform the same series
of experiment but with the three beads slider, in order to test directly the
wall/bead friction properties. At a given $RH$, we indeed observe velocity
strengthening for the sliding of individual steel beads, corroborating
qualitatively the general trend observed on the granular column. But now, we
may go one step further by testing directly the possibility of a
quantitative agreement within the Janssen's model. If we compare these data
to the values of $\mu $ extracted from (\ref{1}) (see fig.3a), assuming the
static value of $K=1.08\pm 0.05$ for all velocities, we observe that the
increase of $\mu $ with $V$ is significantly less important in the case of
the granular column than what is directly measured using the slider device.
Actually, if we suppose a logarithmic increase of $\mu $ with velocity $V$,
i.e. $\mu \sim b\times \log (V)$, we find, at $RH=40\%,$ $b=(2.7\pm
0.2)\times 10^{-2}$ for the slider and $b=(1.2\pm 0.2)\times 10^{-2}$ in the
case of the granular column; in dry conditions, we find $b=(2.4\pm
0.2)\times 10^{-2}$ for the slider and $b=(1.1\pm 0.2)\times 10^{-2}$ in the
case of the granular column. It means that the increase of $F$ with $V$
cannot be entirely attributed to friction effects at the walls, and that the
dynamics may have also an effect on force transmission (i.e. on $K$). In the
framework of a Janssen's analysis it would mean that $K(V)$ would slightly
decrease when velocity increases. Using a simple Hertz law to estimate
contact interactions, we find the depths of penetration of steel beads in
duralumin to be around $\delta \approx 30nm$, whereas in the slider case, we
estimated $\delta \approx 1\mu m$ which is the order of duralumin roughness.
Therefore, it is also possible that contacts are not both in the same
loading regime and then friction laws could be slightly different.
Importantly, we have also found that an increase of humidity has quite a
strong influence on the friction properties (fig.2a). Using the inverted
Janssen's model (eq. (\ref{1})), when assuming the redirection parameter $K$
unchanged by humidity, we recover that the dependence on $RH$\ in the slider
experiment is consistent with the enhancement of the friction forces
measured in the granular column. In a future series of experiments, we will
try to bridge the gap to controlled values of humidity close to $100\%$.

Now let us consider slow driving velocities where the system undergoes a
dynamical instability. A stick-slip motion occurs (see fig.1a) with a narrow
gaussian distribution of slip force amplitudes. In fig.2b, we display the
mean maximum and mean minimum resistance forces (resp. $F_{\max }$ and $%
F_{\min }$) as a function of the driving velocity, for $m=380g$ of beads
(height $H=7.7cm$), and relative humidity $RH=45\pm 3\%$. The mean amplitude
of the slip events $\Delta F=F_{\max }-F_{\min }$ increases strongly when
velocity reaches values as small as $5$ $nm.s^{-1}$. We propose a model
where this enhanced blocking effect can be simply interpreted by an aging
effect of the contacts at the side walls. Friction coefficients of solid on
solid contacts are known to evolve logarithmically with waiting time $t$ 
\cite{Berthoud99}: $\mu _{s}(t)=\mu _{s}^{0}+\beta _{s}\log (t)$. According
to fig.2a, we observe no noticeable variation of $F_{\min }$ with velocity,
for given height and $RH$. Therefore, we will consider in the following $%
F_{\min }$ to be a constant. Starting at the onset of blocking $t=0$, the
force exerted by the force probe during a stick event is $F(t)=F_{\min }+kVt$%
. The time elapsed during a sticking event is: 
\begin{equation}
t_{stick}=%
{\displaystyle{F_{\max }-F_{\min } \over k\times V}}%
.  \label{2}
\end{equation}

The slip occurs when $F(t)$ reaches the maximum force sustainable by the
granular material at time $t$, given by (\ref{1}) with $\epsilon =+1$. The
aging properties of the friction at the wall are included in the time
evolution of the static coefficient of friction $\mu _{s}(t).$ Then we write 
$F(t_{stick})=F_{\max }$, i.e.: 
\begin{eqnarray}
F_{\max } &=&\frac{\varrho g\pi R^{3}}{2K(\mu _{s}^{0}+\beta _{s}\log
(t_{stick}))}  \nonumber \\
&&\times (\exp (2K(\mu _{s}^{0}+\beta _{s}\log (t_{stick}\,\,))\frac{H}{R%
})-1).  \label{3}
\end{eqnarray}

This exponential amplification of the logarithmic aging, due to stress
redirection at the walls, gives an effective power-law : $F_{\max }\sim
t_{stick}^{\,\,\alpha }$, with $\alpha =%
{\displaystyle{2\log (\mbox{e})HK\beta _{s} \over R}}%
$. On fig.3a , for $RH=45\pm 3\%$, we display $\mu _{s}$ extracted from (\ref
{3}) as a function of the time of stick. We assume $K=1.08$ independent both
of the waiting time and the driving velocity. We actually observe a
logarithmic aging for waiting times $\sim 3000s$, with a coefficient $%
\beta _{s}=1.8\times 10^{-2}\pm 2\times 10^{-3}$, value consistent with many
previous reports\cite{Berthoud99}; in the last decade, aging is strongly
increased and we have $\beta _{s}\approx 6\times 10^{-2}$. Note that our
experiment is not a ''clean'' aging experiment since the applied loads and
the shear forces are not constant in time and along the vertical direction.
Furthermore and consistently with the finding of refs \cite{Boquet98} and 
\cite{Crassous99}, we clearly observe that the aging properties are strongly
affected by a variation of the relative humidity $RH$ (see inset of fig.2a).

In conclusion, we investigated the dynamical behavior of a granular column
pushed vertically from the bottom. This model experiment is suited to
understand the rheology of slowly driven granular assemblies in confined
geometries. Overall, the pushing force data are analyzed consistently using
an inverted Janssen's law. At such slow driving velocities we show that, all
the non trivial dynamical properties exhibited by the granular rheology
(including a strong dependence on relative humidity) can be dominantly
attributed to the dynamical properties of solid on solid friction. In
addition, the model seems to indicate the presence of a dynamical structural
effect induced in the bulk at higher driving velocities.

We acknowledge many discussions with Profs R.P Behringer, C.Caroli, T.
Baumberger and the financial support of the CNRS-PICS grant $\#563$.

\newpage
\begin{figure}[h]
\begin{center}
\epsfig{file=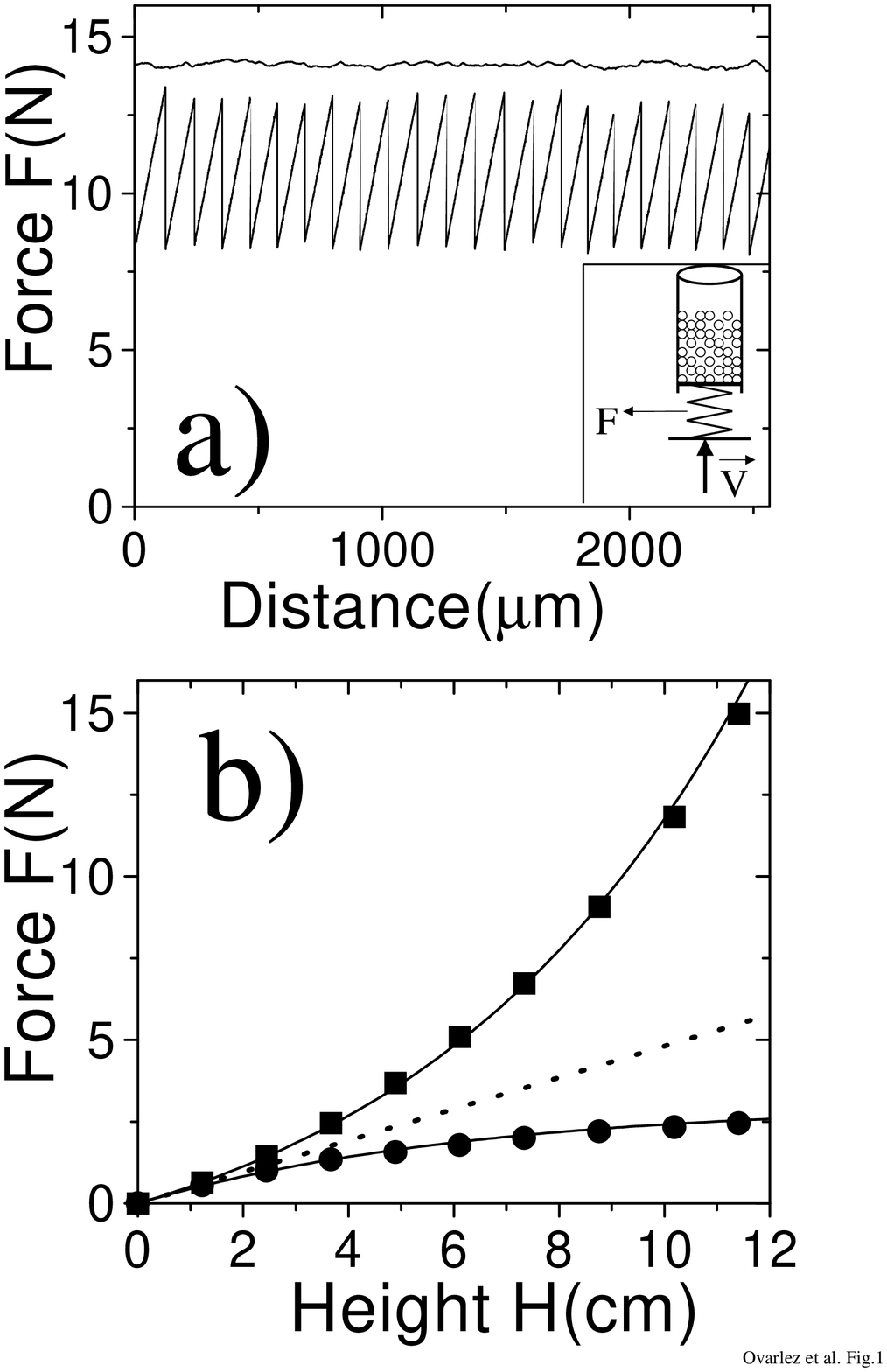,height=19cm}
\caption{a: Resistance force versus the displacement of the stepping motor
for 380 g steel beads in a duralumin cylinder, for 45\% RH and for 2
velocities: $V=30nm.s^{-1}$(stick-slip regime), and $V=100\protect\mu
m.s^{-1}$(steady-sliding regime) shifted by +5N; the inset is a sketch of
the experimental display. b: Mean force in the steady-sliding regime for $%
V_{up}=16\protect\mu m.s^{-1}$ (squares) and for $V_{down}=16\protect\mu
m.s^{-1}$ (triangles) as a fonction of the height of beads; the lines are
the fits according to (1); the dotted line is the hydrostatic curve.}
\label{fig:1}
\end{center}
\end{figure}

\newpage
\begin{figure}[h]
\begin{center}
\epsfig{file=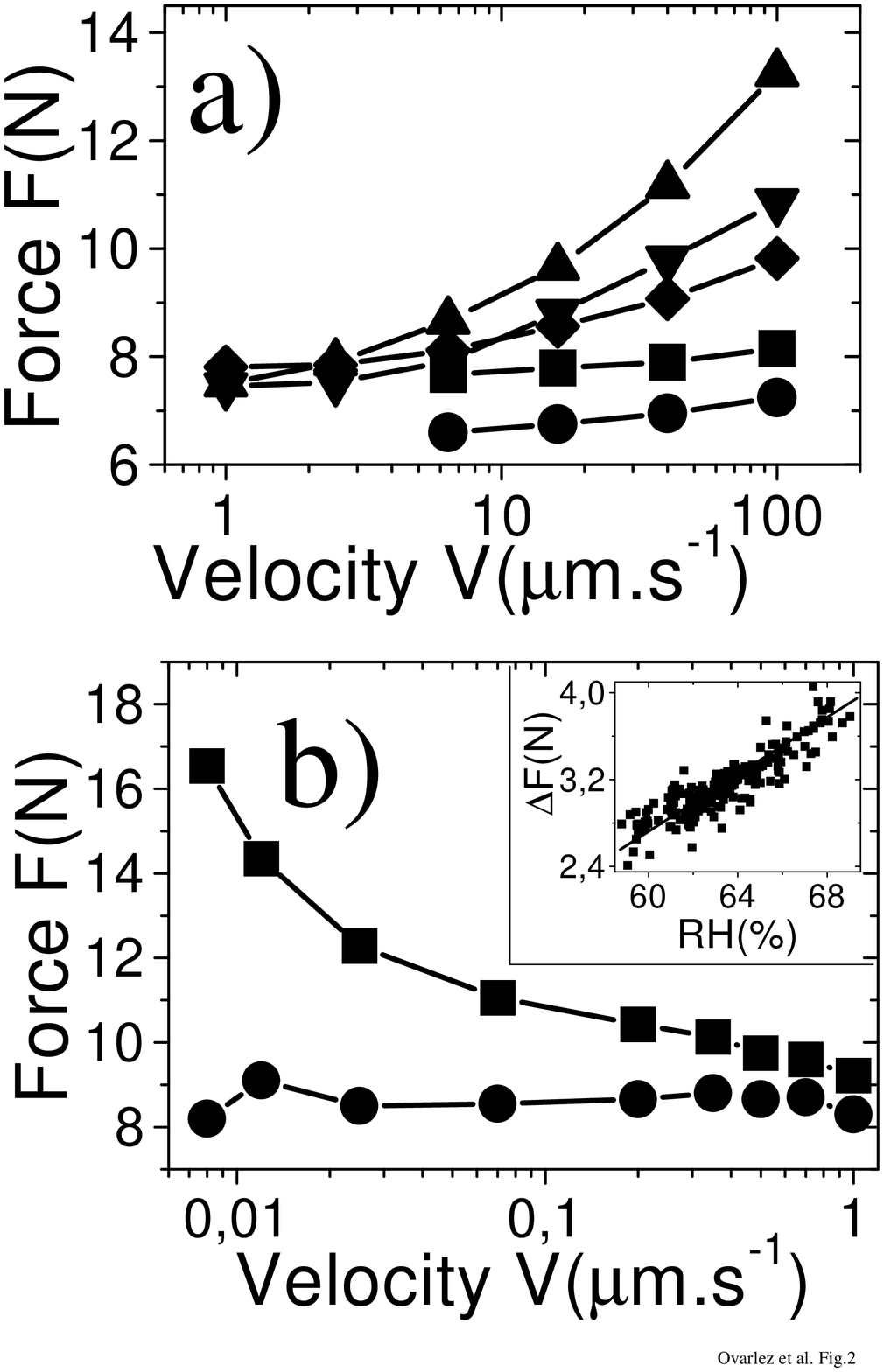,height=19cm}
\caption{ a: Mean force in the steady-sliding regime as a fonction of
velocity for 380g of steel beads in a duralumin cylinder and for several RH (%
$<3\%$ (circles), $40\%$ (squares), $53\%$ (diamonds), $66\%$ (down
triangles), and $72\%$ (up triangles))$.$ b: $F_{\min }$(circles) and $%
F_{\max }$(squares) in the stick-slip regime as a fonction of velocity for
380g of steel beads in a duralumin cylinder and for $RH=48\%$; the inset
shows the variation of $\Delta F=F_{\max }-F_{\min }$ with RH for $%
V=50nm.s^{-1}$. }
\label{fig:2}
\end{center}
\end{figure}

\newpage
\begin{figure}[h]
\begin{center}
\epsfig{file=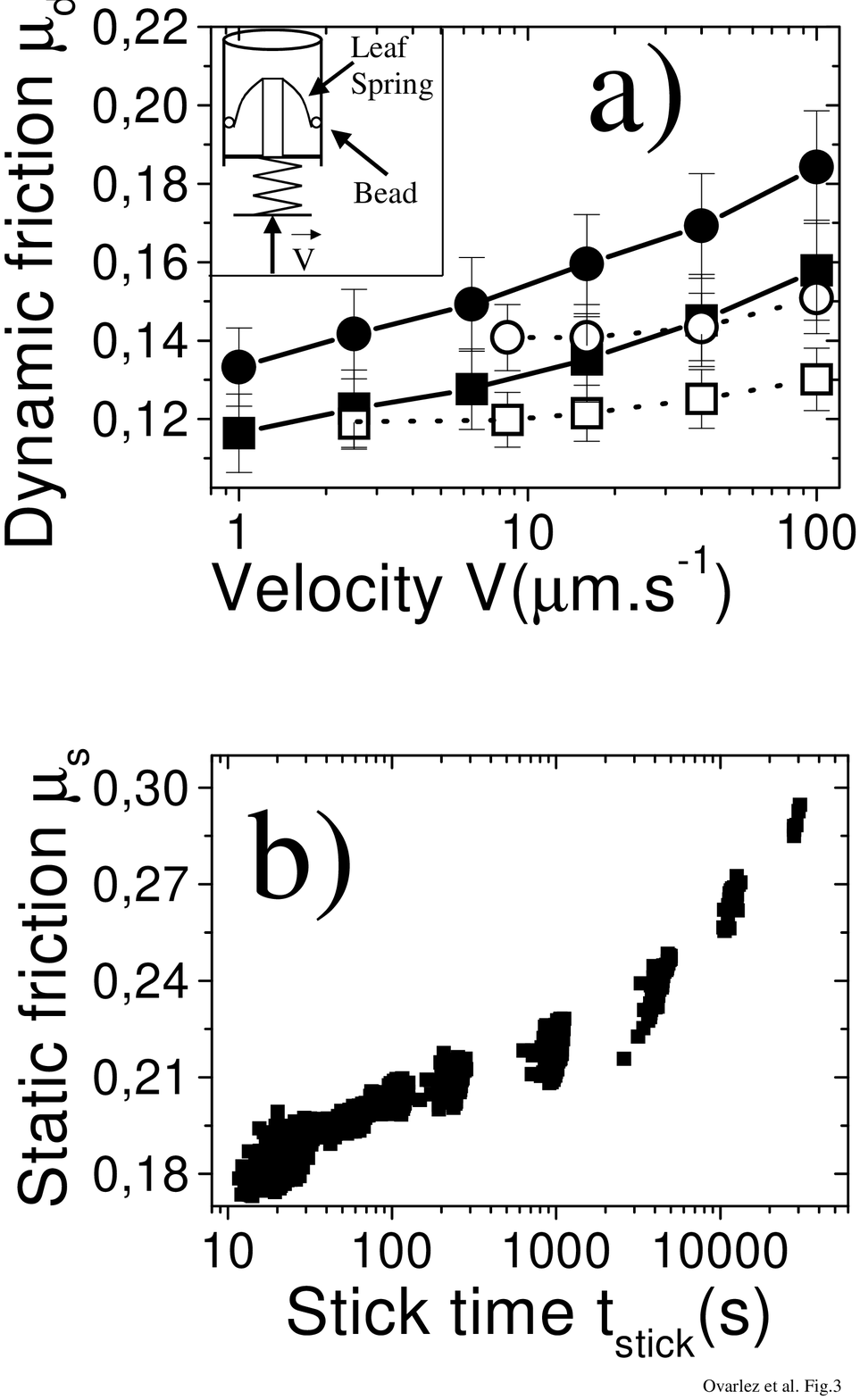,height=19cm}
\caption{ a: Dynamic coefficient of friction as a function of velocity for
the slider (filled symbols) and for the granular column (empty symbols), for 
$RH=40\%$ (circles) and $RH<3\%$ (squares); the inset shows the slider, a
constant normal load is applied on the beads by the way of leaf springs. b:
Static friction coefficient as a function of stick time for 380g of steel
beads in a duralumin cylinder and $RH=45\%$.}
\label{fig:3}
\end{center}
\end{figure}

\end{document}